\documentclass[prd,aps,twocolumn,aps,amsmath,amssymb,nofootinbib,preprintnumbers]
{revtex4}

\usepackage{epsf}
\usepackage{natbib}

\usepackage{graphicx}% Include figure files
\usepackage{dcolumn}% Align table columns on decimal point
\usepackage{bm}% bold math
\usepackage{amsmath}
\usepackage{amsfonts}
\usepackage{subfigure}

\newcommand{\eps}{\epsilon}

\newcommand{\mn}{{\mu\nu}}

\def\be{\begin{equation}}
\def\beq\begin{equation}
\def\ee{\end{equation}}
\def\bea{\begin{eqnarray}}
\def\eea{\end{eqnarray}}

\def\beq{\begin{equation}}
\def\eeq{\end{equation}}
\def\beqa{\begin{eqnarray}}
\def\eeqa{\end{eqnarray}}

\newcommand{\bmat}{\left(\begin{array}}
\newcommand{\emat}{\end{array}\right)}

\newcommand{\co}[3]{\Gamma^{#1}_{#2 #3}}
\newcommand{\chfl}[3]{\genfrac{\{}{\}}{0pt}{}{#1}{#2 #3}}
\newcommand{\half}{\frac{1}{2}}
\newcommand{\inv}[1]{\frac{1}{#1}}
\newcommand{\inta}{\int_U d\Omega}

\newcommand{\bit}{\begin{itemize}}
\newcommand{\eit}{\end{itemize}}
\newcommand{\bnu}{\begin{enumerate}}
\newcommand{\enu}{\end{enumerate}}
\newcommand{\detg}{\sqrt{-g}}
\newcommand{\ba}{\begin{align}}

\begin{document}

\title{Maximal symmetry and metric-affine $f(R)$ gravity}

\author{Tuomas Multam\"aki}
\author{Jaakko Vainio}
\author{Iiro Vilja}
\affiliation{Department of Physics, University of Turku, FIN-20014, Finland}

%\date{May 3, 2006}

\begin{abstract}
%Maximal symmetry is used to reduce the degrees of freedom in metric-affine gravity. 
The affine connection in a space-time with a maximally 
symmetric spatial subspace is derived using the properties of maximally 
symmetric tensors. The number of degrees of freedom in metric-affine gravity is
thereby considerably reduced while the theory allows spatio-temporal torsion and remains 
non-metric. The Ricci tensor and scalar are calculated in terms
of the connection and the field equations derived for the Einstein-Hilbert as wells as 
for $f(R)$ Lagrangians. By considering specific forms of $f(R)$, we demonstrate 
that the resulting Friedmann equations in Palatini formalism without torsion and 
metric-affine formalism with maximal symmetry are in general different
in the presence of matter.
\end{abstract}

\maketitle

\section{Introduction}

Based on the cosmological principle derived from Copernican principle of 
mediocrity and large scale observations, standard cosmology assumes a 
homogeneous and iso\-tro\-pic universe. One finds 
that there are several studies backing this assumption  
(e.g. \cite{hawking73,stoeger1994,Hajian2006,kashlinsky1994}). Although 
the cosmological principle still holds its position as the bedrock of most cosmological 
models, recently the claim for homogeneity has nonetheless been seriously 
contested (e.g. \cite{Alnes2005,enqvist2007,GarciaBellido2008}): intuitively 
one finds this credible as at least on small scales the universe is indeed very 
inhomogeneous.

The idea of homogeneity and isotropy of the universe has been around for a 
long time. Its cosmological implications have been studied thoroughly in the 
context of the metric formalism of the General Relativity (GR). 
Metric-affine formulation of gravity is also an early idea (for its history, see 
e.g. \cite{ferraris82}) based on general concepts of pseudo-Riemannian theory of
manifolds where no {\it a priori} relation between the metric and the connection 
is assumed. However, there have 
been few studies into the effects of homogeneity and isotropy on the independent 
connection in metric-affine 
gravity, probably because the Einstein-Hilbert action does not 
make a distinction between the two formalisms. 

After the initial interest, metric-affine gravity received only 
marginal attention until it flared again in the 1970s \cite{hehl76, hehl77}. 
There were high hopes that metric-affine gravity might lead us closer to quantum 
gravity. Failure to do so lead metric-affine gravity aside once again.
It functioned merely as curiosity until lately the interest in metric-affine 
gravity has grown rapidly since Vollick \cite{vollick2003} argued that it is possible 
to explain the accelerating expansion of the universe without the cosmological 
constant by modifying the Einstein-Hilbert action.

In metric-affine gravity the connection is independent of the metric and has 64 
components which are functions of temporal and spatial coordinates. Intuitively, 
it is clear that by assuming symmetries of the universe, say homogeneity and 
isotropy, the degrees of freedom should decrease. This is indeed well-known
to be true also for the affine connection and the consistent use of symmetry 
principles forms the basis of the present paper. Our aim is to study the general
structure of metric-affine formalism, in the context of $f(R)$ theories of gravity
exploiting the symmetries of homogeneous and 
isotropic universe. % in the context of $f(R)$ theories of gravity. 
More formal studies of $f(R)$ gravity with torsion have also been conducted recently,
see e.g. \cite{Capozziello:2007tj,Capozziello:2008yx} and references therein.

The difference between metric and metric-affine formalisms is manifested by two 
important fundamental features. Torsion is allowed in metric-affine gravity 
unlike in GR (for a review, see \cite{Shapiro:2001rz}). 
The connection can also deviate from GR in non-metricity. 
According to Sotiriou \cite{Sotiriou2006} both can be induced by matter.
However, there is not much experimental evidence to rule out torsion 
(nor non-metricity) or to prove its existence 
\cite{aldrovandi2008,mao2007,hammond2001,Kostelecky:2007kx,Russell:2008ms}. 
This is in part caused 
by the different role it plays in different theories - e.g. in teleparallelism
torsion acts as a force while in GR torsion vanishes by definition and curvature 
geometrizes gravity.

By using symmetry to reduce the degrees of freedom in metric the field equations 
become much more simple. Comparing the results in standard cosmology and results 
in metric-affine formalism it is possible to better see the role which the 
independent connection plays. The present study is organized as follows: In 
section \ref{formi} we devise the general tools needed for the following sections. In
section \ref{homis} we consider a homogeneous and isotropic space and derive the 
independent components of the connection and calculate the Ricci tensor and 
scalar as a function of the found components.
The results of section \ref{homis} are put into use in section \ref{fielde}. 
In the case of Einstein-Hilbert Lagrangian we restrict ourselves to the case of 
empty space and see how the results relate to standard cosmology. Then we 
generalize to $f(R)$ actions and also add matter. In section \ref{conclu} we 
discuss our results.

\section{Symmetry in space-time} \label{formi}

The symmetry of space can be formalized in terms of isometry and form 
invariance. A space is  
form invariant \cite{weinberg1972} under an isometric coordinate 
transformation $x\to \bar x$ if corresponding metric tensors are related by
$\bar g_{\alpha\beta}(y)=g_{\alpha\beta}(y)\ \label{isom}$
for all $y$. %of the metric-affine space-time $M$. 
In the case of infinitesimal 
transformations defined by Killing vectors $\bar x^\mu = x^\mu +X^\mu(x)$ this 
is easily seen to be equivalent with the requirement of vanishing Lie 
derivative $\mathcal L_X g_\mn=0$ \cite{hall2004, gron2007}.
The Lie derivative can be expressed in terms of Levi-Civita connection, 
i.e. Christoffel symbol $\chfl\alpha\mu\nu$ as
\be
\mathcal L_X g_\mn=2\partial_{(\mu}X_{\nu)}-2X_\alpha\chfl\alpha\mu\nu. \label{liekill}
\ee

The affine connection can be most generally written as a sum of a Christoffel 
symbol a torsion part and a non-metricity part \cite{hehl76}. However,
if the connection is metric, i.e. the non-metricity tensor 
$(Q_{\alpha\mn}=-\nabla_\alpha g_\mn)$ vanishes, form invariance can be characterised 
by the Killing equation
\be
\nabla_{(\nu} X_{\mu)}=0 \label{killing}.
\ee
Killing equation still allows for a non-zero torsion tensor \cite{bloomer1977} as
connections of the form
\be
\co\alpha\mu\nu=\chfl\alpha\mu\nu+\half C_\mn{}^\alpha,
\ee
where $C_\mn{}^\alpha$ is antisymmetric in the first two indices, fulfil \eqref{killing} when \eqref{liekill} holds.

For a general tensor we require invariance in an infinitesimal isometric transformation as
for all $y$
\be
T'^{\mn\dots}_{\alpha\beta\dots}(y)=T^{\mn\dots}_{\alpha\beta\dots}(y). \label{forminvar}
\ee
leading to the conditions
\bea
0 & = & \frac{\partial X^\alpha}{\partial x^\mu}T_{\alpha\nu\dots}(x)+
\frac{\partial X^\beta}{\partial x^\nu}T_{\mu\beta\dots}(x)+\nonumber\\
& & \dots+X^\lambda(x)\frac{\partial}{\partial x^\lambda}T_{\mn\dots}(x). \label{lxten}
\eea
In a maximally symmetric space, the requirement that
%The space being maximally symmetric stating that 
the number of independent Killing vectors 
is maximal, i.e. eqs. (\ref {lxten}) are satisfied, strongly
restricts invariant tensors \cite{weinberg1972}.

A scalar in a maximally symmetric space must always be a constant. For higher rank tensors
the invariance equation can be written as
 \be
\delta^\alpha_\mu T^\beta{}_{\nu\dots}+\delta^\alpha_\nu T_\mu{}^\beta{}_{\dots}+\dots=
\delta^\beta_\mu T^\alpha{}_{\nu\dots}+\delta^\beta_\nu T_\mu{}^\alpha{}_{\dots}+\dots 
\label{maxisymten}.
\ee
For our purposes the invariance conditions for tensors of rank $1,\ 2$ and $3$ in a four
dimensional space-time with a maximally symmetric three dimensional subspace are needed.
The first two can be easily found in the literature e.g. \cite{weinberg1972}. For rank three tensor 
the result is seldom calculated explicitly. From here on we use latin indices for the maximally
symmetric subspace while the greek indices refer to four dimensional space-time.

The cases of covariant tensors of rank one and two easily yield that
\begin{subequations}
\ba
A_i&=0 \label{maxvector}\\
B_{ij}&=f g_{ij},
\end{align}
\end{subequations}
where the function $f$ does not depend on the coordinates of the maximally symmetric subspace. 
Applying (\ref{maxisymten}) to a rank three tensor
and contracting indices we get three equations
\begin{subequations}
\ba
(N-1)C_{njk}+C_{jnk}+C_{kjn}&=0 \label{kill61}\\ 
C_{jnk}+(N-1)C_{njk}+C_{nkj}&=0 \label{kill62}\\
C_{njk}+C_{knj}+(N-1)C_{jnk}&=0, \label{kill63}
\end{align}
\end{subequations}
where we have adopted a more general notation with $N$ indicating the dimension of the
maximally symmetric subspace.
From these we obtain two useful conditions for form invariant tensors:
they are invariant under cyclic index permutations,
\be
C_{ k j n}=C_{nkj}\label{kill7},
\ee
and they are antisymmeric in the first two indices, except for $N=3$, since
\be
(N-3)C_{[nj]k}=0.
\ee
From the set of conditions above it follows that all tensors of rank three vanish
unless the maximally symmetric subspace is three dimensional i.e. $N=3$.
As the torsion and non-metricity tensors are rank three, 
they may hence exist only in three dimensional maximally symmetric (sub)spaces (see also \cite{bloomer1977}). 
With $N\neq 3$ the connection is then necessarily the Levi-Civita connection.

\section{Homogeneous and isotropic space} \label{homis}

\subsection{Affine connection}

A metric with a homogeneous and isotropic subspace can be written in spherical
coordinates as \cite{weinberg1972}
\be\label{maxmet}
g_\mn=b^2(t)dt^2-a^2(t)\tilde g_{ij}dx^idx^j,
\ee
where
\be
\tilde g_{ij}=\frac{1}{1-kr^2}dr^2+r^2d\theta^2+r^2\sin^2\theta\, d\phi^2
\ee
is the metric of the spatial part. Usually a rescaling of the time coordinate is performed 
\cite{dirac1975} to remove the function $b(t)$ but at this point we postpone doing this. This
ensures that we can calculate the equations of motion by varying the 
action with respect to $a(t)$ and $b(t)$ instead of varying with respect to the full metric 
tensor.

Taking advantage of the symmetries of space-time, we
require that covariant derivation of a maximally symmetric tensor preserves invariance,
i.e. maximal symmetry. Thereupon we can reduce the number of degrees of freedom in the connection
by utilizing results of the previous section. 
First we consider a maximally symmetric covariant vector $V_\nu$. According to 
\eqref{maxvector} only $V_0\neq 0$ and $V_0=V_0(t)$. Hence
\be
\nabla_0 V_0=\partial_0V_0-\co 000 V_0=b(t) \ \Rightarrow \ \co 000\equiv c_0(t),
\ee
and we see that $\co 000$ depends on time only. 
Moreover
\begin{subequations}
\bea
0 &= & \nabla_0 V_i =\partial_0V_i-\co\alpha 0i V_\alpha \ \Rightarrow \ \co 00i\equiv 0, \\
0 &= & \nabla_i V_0 =\partial_iV_0-\co\alpha i0 V_\alpha \ \Rightarrow \ \co 0i0\equiv 0, \\
f(t)\tilde g_{ij} & =&\nabla_i V_j = \partial_iV_j-\co\alpha ij V_\alpha\nonumber\\
& & \ \Rightarrow \ \co 0ij=-\frac{f(t)}{V_0}\tilde g_{ij}\equiv c_n(t)\tilde g_{ij} \label{colij}.
\eea
\end{subequations}
Keeping in mind that maximally symmetric contravariant vectors only have one nonvanishing component,
one finds that 
\be
\co i0j=c_t(t)\delta^i_j. \label{couij}
\ee

Similar constraints can be derived for rank two tensors,
for example
\be
0=\nabla_0 B_{0i}=\partial_0 B_{0i}-\co\beta 00B_{\beta i}-\co\beta i0 B_{0\beta}=-\co i00 B_{ii} 
\ee
(no sum in the last form), implying that $\co i00=0$. Correspondingly the $0ij$ -component 
gives 
\be 
\co ji0=c_s(t)\delta^j_i.
\ee

The discussion above covers 37 components of the connection reducing them to four independent
components $c_0,\ c_t,\ c_n$ and $c_s$. The last 27 components are found using the results of 
section \ref{formi} in three dimensions. Assuming that the non-metricity tensor vanishes in the 
maximally symmetric subspace the connection can be written as
\be
\co kij
=  \chfl kij+K(t)\epsilon_{ij}{}^k,
\ee
where $\epsilon_{ijk}$ is the three dimensional Levi-Civita symbol. Note that here the 
second term, i.e. the torsion tensor, is invariant under cyclic permutations leaving only one degree 
of freedom. 

Thus the connection preserving maximal symmetry in a three dimensional homogeneous and isotropic 
subspace can be reduced to four spatio-temporal components $c_i(t)$, one component, $K(t)$, 
characterizing spatial torsion and the usual metric Christoffel symbols of a maximally 
symmetric subspace. 
Their usual metric counterparts are 
\begin{subequations}
\ba
c_0&=0 \\
c_s&=c_t=\frac{\dot a}a\\
c_n&=a\dot a\\
K&=0
\end{align}
\end{subequations}
with $b(t)=1$.

\subsection{Ricci tensor and scalar}

The Ricci tensor and curvature scalar are now straightforwardly calculable.
The Ricci tensor is given by \cite{straumann1984}
\be
R_\mn=\partial_\alpha\co\alpha\nu\mu-\partial_\nu\co\alpha\alpha\mu+
\co\beta\nu\mu\co\alpha\alpha\beta-\co\beta\alpha\mu\co\alpha\nu\beta.
\ee
The components $0i$ and $i0$ vanish as they are maximally symmetric 
vectors of rank one in the subspace. The temporal $00$ component reads as
\be
R_{00}=3\big(-\dot c_s+c_0c_s-c_s c_t\big) \label{3symr00}
\ee
and the spatial components can be expressed as
\be
R_{ij}=\tilde R_{ij}+\Big(\dot c_n+c_nc_0+2c_nc_s-c_tc_n\Big)
\tilde g_{ij}+S_{ij}, \label{3symrij}
\ee
where $\tilde R_{ij}$ is the standard Ricci tensor of the spatial part.
Here the last term carries information on spatial torsion,
\be
S_{ij}\equiv\partial_k (K\eps_{ji}{}^k)+K\eps_{ji}{}^k\chfl llk+2K\eps_{l[j}{}^k\chfl l{i]}k
-K^2\eps_{li}{}^k\eps_{jk}{}^l.
\ee

As $S_{ij}$ is antisymmetric and $g_{ij}$ symmetric, contraction of the Ricci tensor
yields
\be\label{cs}
R=-\frac{3}{a^2}\big(2k+\dot c_n+\dot c_s\frac{a^2}{b^2}+2c_nc_s+
C(c_n-c_s\frac{a^2}{b^2})-2K^2\big),
\ee
where we have used the fact that $\tilde R=-6k/a^2$ and denoted $C\equiv c_0-c_t$. 
Note that one can also derive the curvature scalar by using only the torsion tensor 
instead of the connection, as was done in \cite{bloomer1977}.
\bigskip

\section{Field equations} \label{fielde}

\subsection{Einstein-Hilbert action}

Although our goal is to study the results of 
the previous section in a general $f(R)$ model, it is illuminating to consider the  
Einstein-Hilbert action in an empty space. Now the action reads as
\be
S=-\inta \frac{\detg}{2\kappa} R(a,b,c_0,c_t,c_n,\dot c_n,c_s,\dot c_s,K)
\ee
with $R$ given in eq. (\ref{cs}) and $\kappa\equiv 8\pi G$. By variation we obtain the field equations:
\begin{subequations}
\bea
0 & = & 2k+\dot c_n+3\dot c_s\frac{a^2}{b^2}+2c_nc_s+C(c_n-3c_s\frac{a^2}{b^2})\nonumber\\
& & -2K^2 \label{fe1}\\
0 & = & 2k+\dot c_n-\dot c_s\frac{a^2}{b^2}+2c_nc_s+C(c_n+c_s\frac{a^2}{b^2})\nonumber\\
& & -2K^2\label{fe2}\\
0 & = & c_n-\frac{a^2}{b^2}c_s\label{fe3}\\
0 & = & 3\frac{\dot a}{a}-\frac{\dot b}{b}-2\frac{b^2}{a^2}c_n+C 
\label{fe4}\\
0 & = & 2c_s+C-\frac{\dot a}{a}-\frac{\dot b}{b} \label{fe5}\\
0 & = & K \label{fe6}.
\eea
\end{subequations}
Eqs. (\ref{fe1}) and (\ref{fe2}) are exactly the Einstein's equations, albeit written in an unfamiliar
form. Note that $c_0$ and $c_t$ have exactly the same equation of 
motion as they appear in the action only through the combination $C(t)=c_0(t)-c_t(t)$. Therefore here, 
and in the presence of matter as long as it couples only to the metric, there is a spurious degree 
of freedom. 

Combining the results \eqref{fe3} - \eqref{fe6} we find 
\begin{subequations}
\bea
K&=&0, \\
c_s&=&\frac{\dot a}{a}, \\
c_n&=&\frac{\dot aa}{b^2}, \\
C&=&-\frac{\dot a}{a}+\frac{\dot b}{b}.
\eea
\end{subequations}
The first two are exactly the same as in the metric case. The third one coincides also to
the metric case after time rescaling taking $b=1$. The last equation, however, requires more 
analysis. If we assume that torsion vanishes, including spatio-temporal components 
$c_t=c_s$, or that the connection is spatio-temporally metric $\nabla_ig_{0j}=\nabla_ig_{j0}=0$ 
(i.e. $c_t=\frac{b^2}{a^2}c_n=\frac{\dot a}{a}=c_s$) the result is again the same as in 
metric theory. Thus we are left with only one extra degree of freedom which is the
spurious one having physical meaning only if the matter is coupled directly to the connection. 
The maximally symmetric case of Einstein-Hilbert action metric-affine formalism brings 
only one extra degree of freedom when the matter Lagrangian does not depend explicitly on the 
connection i.e. it has no hypermoment \cite{hehl76}. 
This agrees with the idea that hypermoment causes torsion \cite{Sotiriou2006}. 

\subsection{General f(R) Lagrangian}

The analysis in a general $f(R)$ theory with matter follows along similar lines as above.
We assume that the matter Lagrangian $\mathcal L_m$ does not depend explicitly on the 
connection i.e. the hypermoment is zero. 
In this case the gravitational Lagrangian is given by 
$\mathcal L=ba^3f(R(a,b,C,c_n,\dot c_n,c_s,\dot c_s,K))$ and the field equations are now
\begin{subequations}
\bea
2\kappa\frac{T^i{}_i}3 & = & 
f(R)+\frac{2}{a^2}(2k+2c_nc_s+\dot c_n\nonumber\\
& & +Cc_n-2K^2)f'(R)\label{frfe1}\\
2\kappa T^0{}_0 & = & f(R)+\frac{6}{b^2}(\dot c_s-Cc_s)f'(R)\label{frfe2}\\
0 & = & f'(R)\Big(\frac{c_s}{b^2}-\frac{c_n}{a^2}\Big)\\
f''(R)\dot R & = & f'(R)\Big(C+2c_s-\frac{\dot b}{b}-\frac{\dot a}{a}\Big)\label{frfe4}\\
f''(R)\dot R & = & -f'(R)\Big(C-2c_n\frac{b^2}{a^2}-\frac{\dot b}{b}+3\frac{\dot a}{a}\Big)\label{frfe5}\\
0& = & f'(R)K. \label{frfe6}
\eea
\end{subequations}
If $f'(R)\ne 0$, the third and last equations are readily solvable,
\bea
c_s &=& \frac{b^2}{a^2}c_n. \label{frcs},\\
K&=&0.\nonumber
\eea
Summing eqs. \eqref{frfe4} and \eqref{frfe5} and using \eqref{frcs} 
we find
\be
C=\frac{\dot b}{b}-\frac{\dot a}{a} \label{frc}.
\ee
Combining \eqref{frfe1}, \eqref{frfe2}, \eqref{frcs} and \eqref{frc} gives
\be
\frac{b^2}{a^2}c_n^2+k+c_n\Big(\frac{\dot a}{a}-\frac{\dot b}{b}\Big)-\dot c_n=
\frac{\kappa a^2(T^i{}_i-3T^0{}_0)}{6f'(R)}. \label{frab}
\ee
Because the curvature scalar $R$ can be expressed in terms of $c_n$, $a$ and $b$, Eq. 
\eqref{frab} is a nonlinear first order equation for $c_n$. It can be solved, at 
least in principle, for a given $f(R)$.

In the absence of matter summing the first two equations gives the trace equation,
\be
f'(R)R-2f(R)=0,
\ee
implying that empty space is necessarily a space of constant curvature. 
Eqs \eqref{frfe4} and \eqref{frfe5} then yield
\be
c_s=\frac{\dot a}{a}.
\ee
Thus we end up with same components for the connection as for the case of 
Einstein-Hilbert action without matter.
% Again the same spurious degree of freedom, $C=c_0-c_t$, appears. 
We can hence conclude that in a homogeneous and isotropic space without matter,
the metric-affine formalism results in the same equations as metric formalism.
As an easy check shows, adding the cosmological constant leaves the situation unaltered.
Therefore, the possible new effects of metric-affine formalism are due to matter.

With matter that is not coupled to the independent connection, we still get equations 
\eqref{frc}, \eqref{frcs} and \eqref{frab}. The trace equation, however, changes. 
If the matter energy-momentum tensor is of perfect fluid form we have
\be
f'(R)R-2f(R)=\kappa(3p-\rho). \label{tracemat}
\ee
Here we note that in the special case of radiation filled universe the right 
hand side vanishes and once again we reproduce the results of metric formalism. 
Moreover, if the hypermoment were present all the aforementioned equations would change. 
Even the simple \eqref{frfe6} would become non-trivial and giving $K\propto (a^3f'(R))^{-1}$. 
As the nature of the gravitation-matter coupling is not completely clear
even this approach has some potential interest.

Although a radiation dominated universe reproduces the metric cosmology, this is not
a general property. For example, if we choose $f(R)=R+\lambda R^2$, with $\lambda$ some small 
constant, and examine a non-relativistic matter filled universe, the trace equation 
\eqref{tracemat} yields
\be
R=\kappa\rho=\frac{\kappa\rho_0}{a^3}, \label{rhor}
\ee
where $\rho_0$ is a constant and we have rescaled time so that $b=1$. 
From equations \eqref{frcs}, \eqref{frc} and \eqref{frfe4} we then get
\be
c_n=\frac{a\dot a(a^3-\kappa\rho_0\lambda)}{2\kappa\rho_0\lambda+a^3}. \label{r2cn}
\ee
Clearly we need $\rho_0=0$ in order to reproduce $c_n=\dot aa$ (i.e. the metric solution), 
leaving empty space as the only possibility. If, however, we allow for non-Levi-Civita connections 
there are other possibilities. Inserting Eq. (\ref{r2cn}) into \eqref{frfe2} and 
\eqref{rhor} we can eliminate $\ddot a$ to obtain an effective Friedmann equation
\be\label{fried}
H^2=-\frac{(2\kappa\lambda+a^3)(\half\kappa^2\lambda\rho_0^2+6\kappa\lambda ka-
\kappa\rho_0a^3+3ka^4)}{3a^3(a^3-\kappa\lambda\rho_0)^2}. 
\ee
If we expand this equation in $\lambda$, the result is the more intuitive
\be
H^2=\frac {\kappa\rho_0}{3a^3}-\frac k{a^2}+\Big(\frac{7\kappa^2\rho_0^2}{6a^6}-
\frac{6\kappa\rho_0k}{a^5}\Big)\lambda+\mathcal O(\lambda^2). \label{h2r+r2}
\ee
The limit $\lambda\rightarrow 0$ coincides with standard cosmology as expected. 
Note, that the correction $\propto a^{-6}$ can be created also by adding
non-metric matter coupling, i.e. hypermoment, as in \cite{bloomer1977}, but
here it is created solely by the form of the gravitational Lagrangian. 
Comparing Eq. \eqref{fried} to the results in the Palatini formalism 
(without torsion) \cite{vollick2003,Koivisto2005,kainulainen2006} we find that 
they agree.

This raises the question, whether our maximally symmetric approach generally 
coincides with the results in the more commonly considered Palatini formalism.
In order to answer this question, we consider a toy model where 
the Lagrangian is of the form $f(R)=R^n$.
Following the procedure above results in an effective Friedmann equation
\be
H^2=-\frac{4n^2k}{(n-3)^2a^2}-\frac{2n(n+1)}{3(n-3)^2}A^{\inv n},
\ee
where $A=\frac{\kappa\rho_0}{(n-2)a^3}$. The corresponding equation in the Palatini formalism 
reads as
\be
H^2=\frac{2n\big((1-n)A^{\inv n}a^3+2\kappa\rho_0 A^{\frac{1-n}n}-6nka\big)}{3a^3\big((7n+6)n-9\big)}.
\ee
Hence, we see that the conincidence in the $\lambda R^2$ model was an exception: the maximally 
symmetric formalism and the Palatini formalism in general lead to different dynamical equations.
The difference is pronounced in the case of $n=3$, where the Palatini formalism is well-behaved but
here we find that our approach is singular in the sense
that no Friedmann equation can be derived. Note, that there is also singularity at 
$n=2$ in both cases as the trace equation for $f(R)=R^2$ holds only in empty space.
Our result should be compared with the result of \cite{Capozziello:2007tj} where it was found that
metric-affine formalism with torsion only, i.e. with fully vanishing non-metricity does
coincide with Palatini formalism.

\section{Conclusions} \label{conclu}

In this paper we have studied a homogeneous and isotropic space with a maximally symmetric formalism 
in $f(R)$ theories of gravity.
The effects of homogeneity and isotropy in the standard Einstein-Hilbert case has been discussed 
before \cite{bloomer1977} but here we have shown that even in more general $f(R)$ theories, 
only one spurious extra degree of freedom appears in empty space.

Interesting possibilities begin to emerge, when one includes matter in the system. 
In the case of the Einstein-Hilbert action, the addition of matter without hypermoment 
does not change the solutions of the field equations from those of metric formalism. 
New types of solutions appear only if the matter Lagrangian has an explicit dependence on the connection 
\cite{bloomer1977}, in which case the connection is even less determined for general $f(R)$ Lagrangians.
These results are in accordance with those of \cite{Sotiriou2006} where it was argued that torsion 
is caused by the antisymmetric part of the hypermoment.

However, even for ordinary matter (i.e. no hypermoment), the construction of the Friedmann equations 
reveal that the maximally symmetric formalism is dynamically different from the corresponding
Palatini formalism although they may coincide in some special cases. This appears to be a consequence
of inclusion of spatio-temporal non-metricity. Indeed
the difference between the two formalisms is due to the fact that 
in the Palatini formalism torsion is assumed to vanish {\it a priori} whereas here 
only spatial non-metricity is 
assumed to vanish. Therefore the degrees of freedom in these two approaches are dissimilar resulting 
in a differently constrained system. Physically it is unclear which approach one should adopt.
As there is almost no evidence for
torsion, the usual pick would be Palatini formalism. Metric-affine formalism, however, 
is more general and is based on the explicit use of the cosmological principle.   

In all cases a spurious degree of freedom which has little or no physical meaning remains. 
It emerges because two components of the connection appear only as a certain combination in the Lagrangian.
As they affect the physics of the universe only via this combination, their geometrical 
interpretation can be found if there are non-metric matter couplings present.

The cosmological consequences of the maximally symmetric formalism is an interesting possible
direction of studies as well as generalization to spherically symmetric systems. Both are 
likely to give at least some constrains for a given $f(R)$ theory. Furthermore, although isotropy is 
commonly accepted there have been numerous articles investigating the possibility of an inhomogeneous 
universe \cite{Alnes2005,enqvist2007,GarciaBellido2008,alnes2006, garfinkle2006},
motivating further study of the the connection in an inhomogeneous and isotropic space. 
These results could be used to ease the usage of metric-affine formalism in spherically
symmetric universes.

%\acknowledgments

\addcontentsline{toc}{chapter}{Viitteet}
\bibliography{refs}

\end{document}